\documentclass[aps,pra,amssymb,superscriptaddress,footinbib,showpacs,twocolumn,longbibliography]{revtex4-1} 
\usepackage{amsmath}
\usepackage{amsfonts}
\usepackage{graphicx}
\usepackage{epsfig}
\usepackage{subfigure}
\usepackage{bbm}
\usepackage{psfrag}
\usepackage{pstool}
\usepackage{mathtools}

\newcommand{\ba}{\begin{align}}
\newcommand{\ea}{\end{align}}
\newcommand{\be}{\begin{equation}}
\newcommand{\ee}{\end{equation}}
\newcommand{\bea}{\begin{eqnarray}}
\newcommand{\eea}{\end{eqnarray}}

\newcommand{\lr}{\left(}
\newcommand{\rr}{\right)}

\newcommand{\leff}{\delta L_{\text{eff}}}
\DeclareMathOperator\arctanh{arctanh}

\begin{document}
\title{Simulating moving cavities in superconducting circuits}
\author{Stefano Bosco}
\thanks{These two authors contributed equally to the manuscript.}
\affiliation{
Institute for Quantum Information, RWTH Aachen University,                                
  D-52056
  Aachen,                              
  Germany
}

\affiliation{
Peter Gr\"{u}nberg Institute, Theoretical Nanoelectronics, Forschungszentrum J\"{u}lich, D-52425 J\"{u}lich, Germany
}
\author{Joel Lindkvist}
\thanks{These two authors contributed equally to the manuscript.}
\affiliation{Microtechnology and Nanoscience, MC2, Chalmers University of Technology, S-41296 G\"oteborg, Sweden}
\author{G\"oran Johansson}
\affiliation{Microtechnology and Nanoscience, MC2, Chalmers University of Technology, S-41296 G\"oteborg, Sweden}

\begin{abstract}
We theoretically investigate the simulation of moving cavities in a superconducting circuit setup. In particular, we consider a recently proposed experimental scenario where the phase of the cavity field is used as a moving clock. By computing the error made when simulating the cavity trajectory with SQUIDs, we identify parameter regimes where the correspondence holds, and where time dilation, as well as corrections due to clock size and particle creation coefficients, are observable. These findings may serve as a guideline when performing experiments on simulation of moving cavities in superconducting circuits.
\end{abstract}
\maketitle
\section{Introduction}
Besides its potential in quantum information processing \cite{anton2017,wendin2017}, the field of superconducting circuits has in recent years emerged as a platform for simulating relativistic physics. 
Several experimental scenarios have been proposed, all based on the tunability of the superconducting quantum interference device (SQUID). These include relativistic motion of qubits \cite{relativisticqubits} for observation of the cavity-enhanced Unruh effect \cite{unruhincavity}, and Hawking radiation \cite{hawking1,hawking2} in an analogue Schwarzchild spacetime \cite{analoguehawking}.

To date, the most successful simulation of relativistic physics in superconducting circuits has been the realization of tunable boundary conditions for the electromagnetic field mimicking those of moving mirrors. In 1970, it was predicted that oscillating boundary conditions create photons from the vacuum. This phenomenon, known as the dynamical Casimir effect (DCE), was first described for a field in a tunable-length cavity \cite{moore70} and later generalized to an oscillating mirror in open space \cite{fulling76}. Due to the relativistic speeds involved, however, it has proven difficult to observe the DCE using mechanical mirrors. To circumvent the problem, in \cite{dcepra}, it was proposed to measure DCE radiation in a superconducting circuit setup.  The experiment was later successfully carried out in \cite{dcenature}, leading to the first-ever observation of the DCE.
\newline
\\
In relativistic quantum information (RQI) \cite{rqireview,ivettereview}, the effects of relativity in quantum information processing are investigated. While many earlier works in RQI concerned global modes of the electromagnetic field \cite{alsing03,alicefalls,adesso07}, it is of interest both conceptually and practically to study also localised systems. One approach to achieve this is to confine the field in a rigid cavity moving at relativistic speed.
Using the rigid cavity model, the effects of relativistic motion on quantum properties such as entanglement have been theoretically studied \cite{downes11,voyageto12,motiongenerates12,quantumgates12}. However, for the same reasons that the DCE eluded experimental verification for many years, these results are difficult to verify in experiments with mechanical cavities. In \cite{teleportationfriis}, it was first suggested that the setup used to observe the DCE can be extended in order to simulate a relativistically moving cavity, making superconducting circuits a possible test ground for RQI. 
\newline
\\
A specific experimental proposal involving a moving cavity in superconducting circuits was made in \cite{twinparadox}. There, the idea was to observe time dilation by using the phase of a cavity field mode as a clock, and the same clock model was studied in \cite{lindkvist6}. In \cite{twinparadox}, only an ideal situation was examined, and estimations were made to find possible parameter regimes for an experiment. In this paper, we instead perform the calculations in more detail from the circuit perspective. We can thereby examine how good the analogy between the circuit setup and the moving cavity is in different regimes.

The paper is organized as follows. In section \ref{sec:dirichlet}, we review the description of rigid cavities and the clock model used in the previous papers. In section \ref{sec:squidbc}, we introduce the actual boundary conditions imposed by the SQUIDs in the circuit setup and in section \ref{sec:robin} we outline how to compute transformations of the cavity state using these boundary conditions. In section \ref{sec:simulation}, we calculate the phase shift of the cavity clock mode using the SQUID boundary conditions and investigate how well the correspondence works in different parameter regimes. In section \ref{sec:harmonicflux}, we discuss how to achieve shorter trajectory times for the cavity and in section \ref{sec:resonance}, particle creation resonances are investigated. Finally, in section \ref{sec:conclusions} we summarize and conclude our work.

\section{Rigid Dirichlet cavities}
\label{sec:dirichlet}
We consider a rigid cavity with perfectly conducting boundaries and an electromagnetic field confined inside. The rigidity ensures that the proper length $L$ of the cavity is preserved throughout the experiment. When the motion is constrained to one dimension, the cavity can be modeled by imposing Dirichlet boundary conditions for a massless Klein-Gordon scalar field in 1+1-dimensional spacetime. For an inertial observer in Minkowski spacetime, the Klein-Gordon field $\phi$ obeys the wave equation
\be
\lr\partial^2_{t}-c^2\partial_x^2\rr\phi=0,
\label{eq:kgminkowski}
\ee
where $c$ is the speed of light in the medium. Imposing Dirichlet boundary conditions at $x=x_l$ and $x=x_r$, with $L=x_r-x_l$, we obtain the following complete set of orthonormal mode functions
\be
u_n(t,x)=\frac{1}{\sqrt{\pi n}}\sin{\lr \frac{\omega_n}{c}(x-x_l)\rr}e^{-i\omega_nt},
\label{eq:dirichletmodefunctions}
\ee
with $\omega_n=\pi n c/L$, $n\in \mathbb{N}$.

Stepping to a quantum description, the field operator is expanded as
\be
\phi(t,x)=\sum_n\lr a_nu_n(t,x)+a_n^{\dagger}u^*(t,x)\rr,
\label{eq:minkowskifield}
\ee
where the annihilation and creation operators satisfy the canonical commutation relations $[a_m,a_n^{\dagger}]=\delta_{mn}$.
\newline
\\
Uniformly accelerated observers follow hyperbolic trajectories through spacetime. These worldlines can be conveniently parametrized using the Rindler coordinates $(\eta,\xi)$, defined by
\bea
x&=&\frac{c^2}{\alpha}e^{\alpha\xi/c^2}\cosh{\lr \alpha\eta/c\rr},\\
t&=&\frac{c}{\alpha}e^{\alpha\xi/c^2}\sinh{\lr \alpha\eta/c\rr},
\eea
where $\alpha$ is the constant proper acceleration of an observer at Rindler position $\xi=0$. Observers at different Rindler locations are moving along hyperbolic trajectories with different proper accelerations.
Expressed in Rindler coordinates, the wave equation \eqref{eq:kgminkowski} becomes
\be
\lr\partial^2_{\eta}-c^2\partial_{\xi}^2\rr\phi=0.\label{rindlerwave}
\ee
To describe a uniformly accelerated rigid cavity of proper length $L$, we impose Dirichlet boundary conditions at $\xi=\xi_l$ and $\xi=\xi_r$, with $L'=\xi_r-\xi_l$.  Here, $L'$ is the Rindler length of the cavity, related to the proper length $L$ by
\be
L'=\frac{\arctanh{\lr h/2\rr}}{(h/2)}L,
\ee
where we have defined the dimensionless parameter
\be
h\equiv aL/c^2,\label{hdef}
\ee
with $a$ being the proper acceleration of an observer in the center of the cavity.
Note that the rigidity constraint implies that each point in the cavity must move with a different proper acceleration. It also implies $h<2$, since the proper acceleration of the rear end of the cavity approaches infinity for $h\rightarrow 2$.

A complete set of orthonormal mode functions satisfying \eqref{rindlerwave} with Dirichlet boundary conditions is given by
\be
v_m(\eta,\xi)=\frac{1}{\sqrt{\pi m}}\sin{\lr\frac{\Omega_m}{c}(\xi-\xi_l)\rr}e^{-i\Omega_m\eta},\label{rindlerfunction}
\ee
with $\Omega_m=\pi m c/L'$, $m\in \mathbb{N}$.
Similarly to \eqref{eq:minkowskifield}, the field operator can be expanded as
\be
\phi(\eta,\xi)=\sum_m\lr b_mv_m(\eta,\xi)+b_n^{\dagger}v^*(\eta,\xi)\rr,
\label{eq:rindlerfield}
\ee
with $[b_m,b_n^{\dagger}]=\delta_{mn}$.
\newline
\\
To describe the time evolution of the cavity field during motion, we work in the Heisenberg picture. During inertial motion, the field undergoes free Minkowski time evolution and the annihilation operators $a_n$ pick up phase factors determined by $\omega_n$. Likewise, under uniform acceleration, the field evolves freely in Rindler time $\eta$ and the operators $b_m$ acquire phase factors determined by $\Omega_m$.

Consider now the case when a rigid cavity in inertial motion suddenly begins to accelerate uniformly at time $t=0$.
At this instant, the field can be expanded in both the Minkowski and the Rindler basis. Equating \eqref{eq:minkowskifield} and \eqref{eq:rindlerfield} results in the Bogoliubov transformation
\be
b_m=\sum_n\lr \alpha_{mn}^*a_n-\beta_{mn}^*a_n^{\dagger}\rr,
\label{eq:bogotransf}
\ee
where the coefficients can be expressed in terms of the Klein-Gordon inner product \cite{birrell1984} as $\alpha_{mn}=(v_m,u_n)$ and $\beta_{mn}=(-v_m,u_n^*)$. These coefficients are functions of the parameter $h$, defined in \eqref{hdef}.

The evolution of the cavity field during a trip consisting of segments of inertial motion and uniform acceleration is determined by composing transformations of the form \eqref{eq:bogotransf} and their inverses, with free Minkowski and Rindler time evolution in between. The resulting transformation has the same form as \eqref{eq:bogotransf}, where $a_n$ and $b_m$ are now related to the modes in the cavity before and after the trip, respectively.
\newline
\\
The relativistically rigid cavity described above provides a localized system in quantum field theory. For this reason, it has been theoretically studied in the context of relativistic quantum information to investigate the effects of motion on quantum properties such as entanglement \cite{downes11,voyageto12,motiongenerates12,quantumgates12}. In \cite{twinparadox}, it was suggested that the same system can be used as a fundamental physical model of a moving clock. Since the phase shift of a coherent state is proportional to the elapsed proper time, the phase of the state can be used as a clock pointer. In this way, time dilation is detected by comparing the phase shift to that of a static reference cavity.
From the Bogoliubov coefficients in \eqref{eq:bogotransf}, the phase shift $\Delta\theta$ of the first cavity mode is determined by
\be
\tan{\Delta\theta}=-\frac{\text{Im}\lr\alpha_{11}-\beta_{11}\rr}{\text{Re}\lr\alpha_{11}-\beta_{11}\rr}.
\ee
In the following, unless otherwise stated, we always compute this phase shift relative to the corresponding phase shift in a static Dirichlet cavity.

\begin{figure}[t]
\centering
\includegraphics[width=0.9\columnwidth]{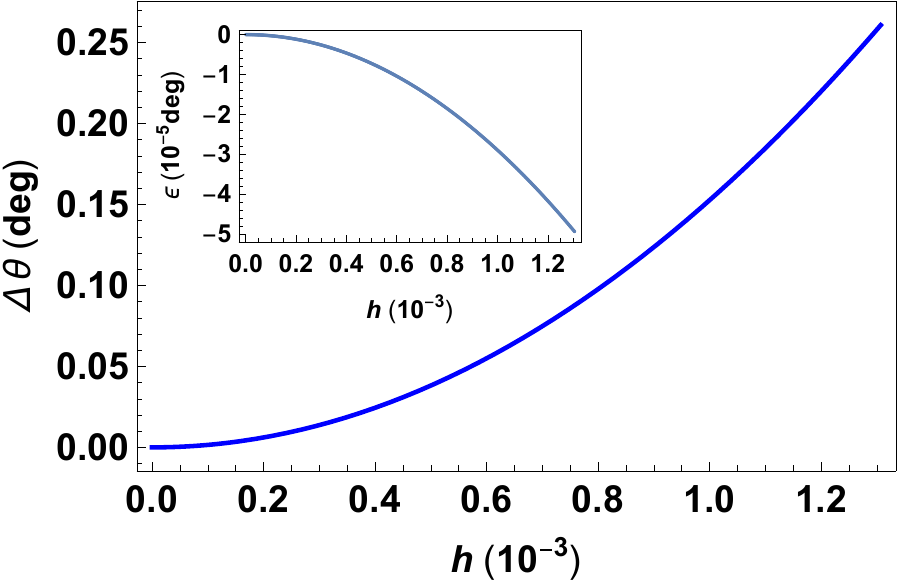}
\caption{Phase shift of a single trajectory of the type considered in figure 3 in \cite{twinparadox} (see the Appendix), computed using the full Bogoliubov coefficients. The inset shows the absolute error $\epsilon$ made when computing the same phase shift using second-order Bogoliubov coefficients.}
\label{fig:MRperturbativeComparison}
\end{figure}

The higher moments of the transformed field were calculated in \cite{lindkvist6}, making it possible to compute the variance of the phase and thus the clock precision.
The clock described above has a finite extension and is based on quantum field theory, leading to deviations from the ideal clock formula \cite{dragan15}. 
\newline
\\
In \cite{twinparadox}, a class of simple round trip trajectories consisting of four hyperbolic segments with the same proper acceleration was studied. These trajectories are specified by the proper length $L$ of the cavity, the proper acceleration $a$ in the middle of the cavity and the lab frame acceleration time $t_a$ for each segment, and the total displacement of the cavity is given by
\be
d_{\text{cav}}=2\sqrt{c^2t_a^2+c^4/a^2}-\frac{2c^2}{a}.\label{cavdisp}
\ee
The explicit trajectory functions are given in Appendix \ref{sec:appendix}.
In \cite{twinparadox}, the phase shift accumulated over these trajectories was determined by computing the Bogoliubov coefficients analytically to second order in $h$. In this work, we compute the coefficients numerically for arbitrary allowed values of $h$. In this way, the validity of previous approximations can be addressed and different parameter regimes can be explored. Moreover, as discussed in section \ref{sec:resonance}, our numerical analysis allows us to easily compose transformations to investigate resonant effects. Figure \ref{fig:MRperturbativeComparison} shows the phase shift considered in \cite{twinparadox} in the regime suggested for the experiment. In the inset plot, we clearly see that the error in using the perturbative approach is small.

\section{SQUID boundary conditions}
\label{sec:squidbc}
The idea to use a SQUID in a superconducting coplanar waveguide to implement a time-dependent Dirichlet boundary condition was first realized in a demonstration of the dynamical Casimir effect \cite{dcepra,dcenature}. In \cite{teleportationfriis}, it was later suggested that a moving cavity can be simulated in a similar manner and a more detailed experimental proposal involving this type of cavities was made in \cite{twinparadox}.

A superconducting coplanar waveguide (CPW) supports a 1+1-dimensional massless Klein-Gordon field $\phi(t,x)$ propagating at speed $c=1/\sqrt{L_0C_0}$, where $L_0$ and $C_0$ are the inductance and capacitance per unit length of the waveguide. By terminating the waveguide by one DC SQUID at $x=0$ and another at $x=L_{\text{cav}}$, the boundary conditions of the field can be written in the following way \cite{dcepra}
\bea
\phi(t,0)-\delta L_{\text{eff}}(\Phi_l)\partial_x\phi(t,x)|_{x=0}&=&0,\label{eq:robinbc1}\\
\phi(t,L_{\text{cav}})+\delta L_{\text{eff}}(\Phi_r)\partial_x\phi(t,x)|_{x=L_{\text{cav}}}&=&0,\label{eq:robinbc2}
\eea
with
\be
\delta L_{\text{eff}}\lr\Phi\rr=\frac{\Phi_0}{2\pi}\frac{1}{2L_0I_c\left|\cos{\lr\pi\Phi/\Phi_0\rr}\right|}.\label{eq:effectivelength}
\ee
Here, $I_c$ is the critical current of the Josephson junctions in the SQUID, $\Phi_0=h/2e$ the magnetic flux quantum and $\Phi$ the external magnetic flux threading the SQUID loop; the index $l(r)$ labels the left (right) end of the CPW. It has also been assumed that the plasma frequency of each SQUID is far greater than any other relevant frequencies. The parameter $\delta L_{\text{eff}}(\Phi)$ can be dynamically tuned by modulating the external magnetic flux.
\begin{figure}[t]
\centering
\includegraphics[width=0.9\columnwidth]{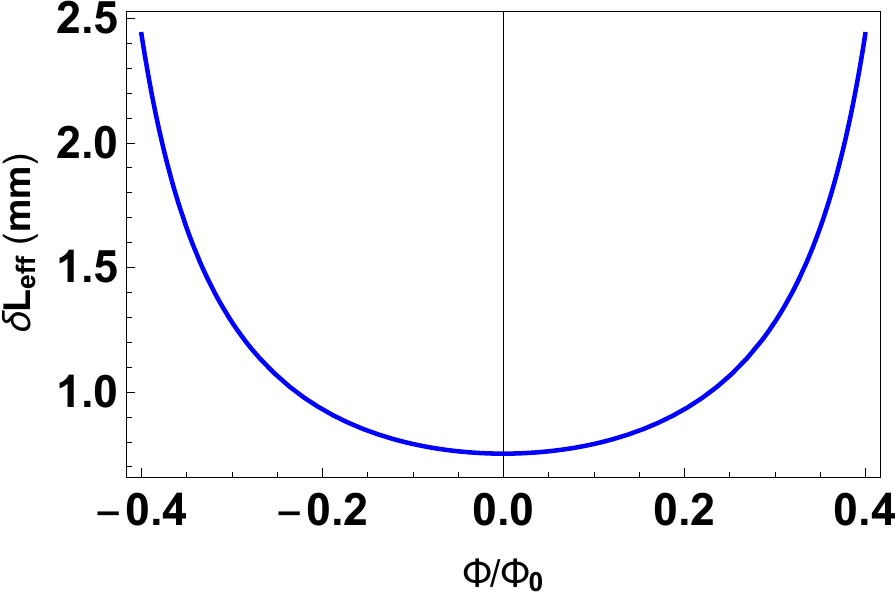}
\caption{Extra effective length $\leff(\Phi)$ at the end of a coplanar waveguide terminated by a DC SQUID, as a function of the external magnetic flux $\Phi$ through the SQUID. The physical parameters of the SQUID and the waveguide are chosen as in \cite{lindkvist6}, $L_0=0.44 \mu \mathrm{F/m}$ and $I_c=0.5 \mu \mathrm{A}$.}
\label{fig:effectivelength}
\end{figure}

Equations \eqref{eq:robinbc1} and \eqref{eq:robinbc2} are known as Robin boundary conditions and for wavelengths $\lambda$ that satisfy $\lambda\gg \delta L_{\text{eff}}(\Phi_l), \delta L_{\text{eff}}(\Phi_r)$, they are approximately equal to
\bea
\phi(t,-\delta L_{\text{eff}}(\Phi_l)=0,\label{effectivedirichlet1}\\
\phi(t,L_{\text{cav}}+\delta L_{\text{eff}}(\Phi_r))=0.\label{effectivedirichlet2}
\eea
These are Dirichlet boundary conditions at effective positions $x=-\delta L_{\text{eff}}(\Phi_l)$ and $x=L_{\text{cav}}+\delta L_{\text{eff}}(\Phi_r)$. Thus, while the physical length of the cavity is $L_{\text{cav}}$, its effective length in the lab frame is equal to $L_{\text{cav}}+\delta L_{\text{eff}}(\Phi_l)+\delta L_{\text{eff}}(\Phi_r)$. Figure \ref{fig:effectivelength} shows the extra effective length \eqref{eq:effectivelength} at one end of the CPW as a function of the external flux.  By dynamically tuning the SQUID fluxes $\Phi_l$ and $\Phi_r$, the two boundary conditions can be modified independently of one another in a way that can simulate two moving mirrors. To simulate cavity motion to the right, one would have to decrease (increase) $\delta L_{\text{eff}}\lr\Phi_l\rr (\delta L_{\text{eff}}\lr\Phi_r\rr)$, and the other way around for leftward motion.
\newline
\\
In \cite{twinparadox}, the setup described above was proposed in order to simulate a rigidly moving cavity and detect time dilation.
The calculations, however, were made only for ideal rigid Dirichlet cavities. In this paper, we perform the analysis using the full Robin boundary conditions \eqref{eq:robinbc1}-\eqref{eq:robinbc2}. Our goal is to analyze the limitations of the analogy between the SQUID and a moving mirror in different parameter regimes.

\section{Time-dependent Robin boundary conditions}
\label{sec:robin}
\begin{figure*}[t]
\centering
\subfigure{
	\includegraphics[width=0.9\columnwidth]{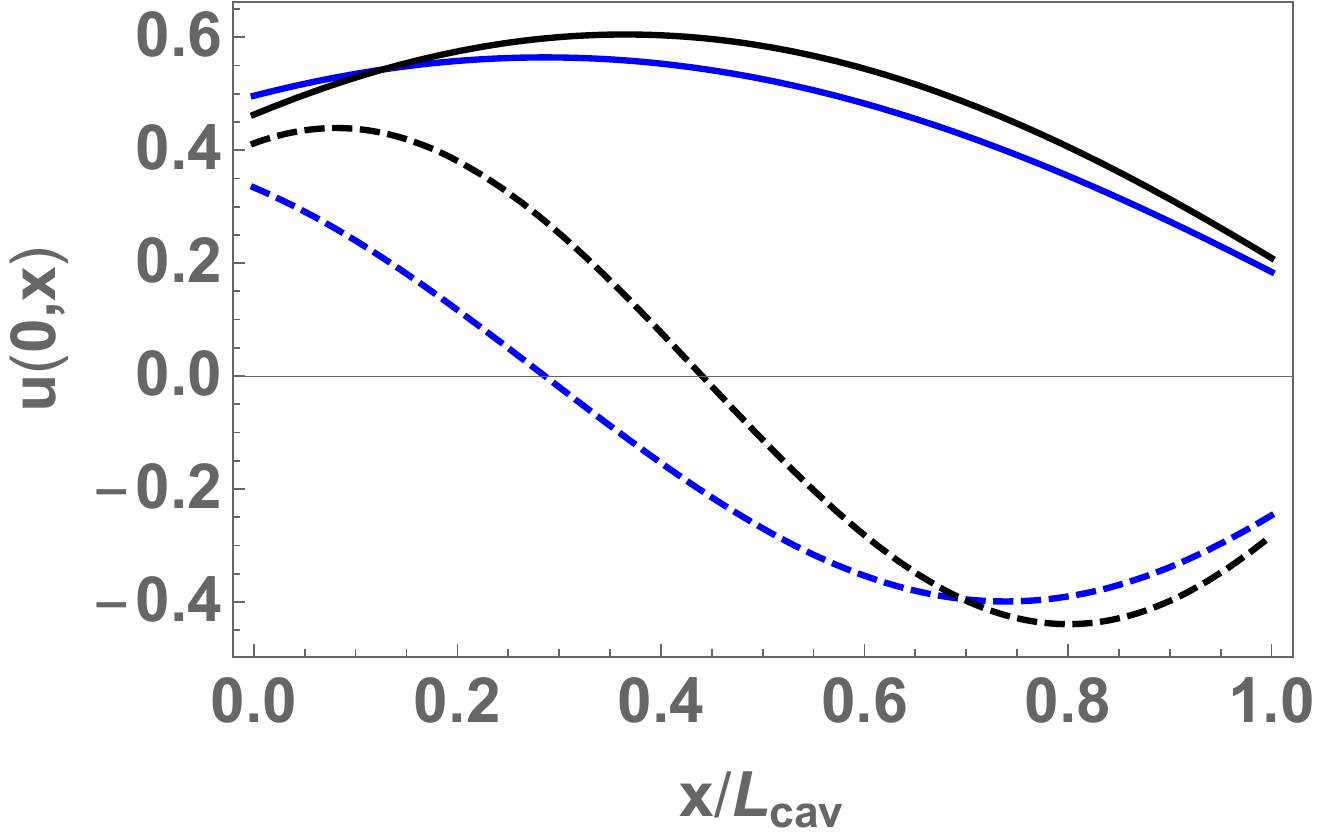}
}
\subfigure{
	\includegraphics[width=0.9\columnwidth]{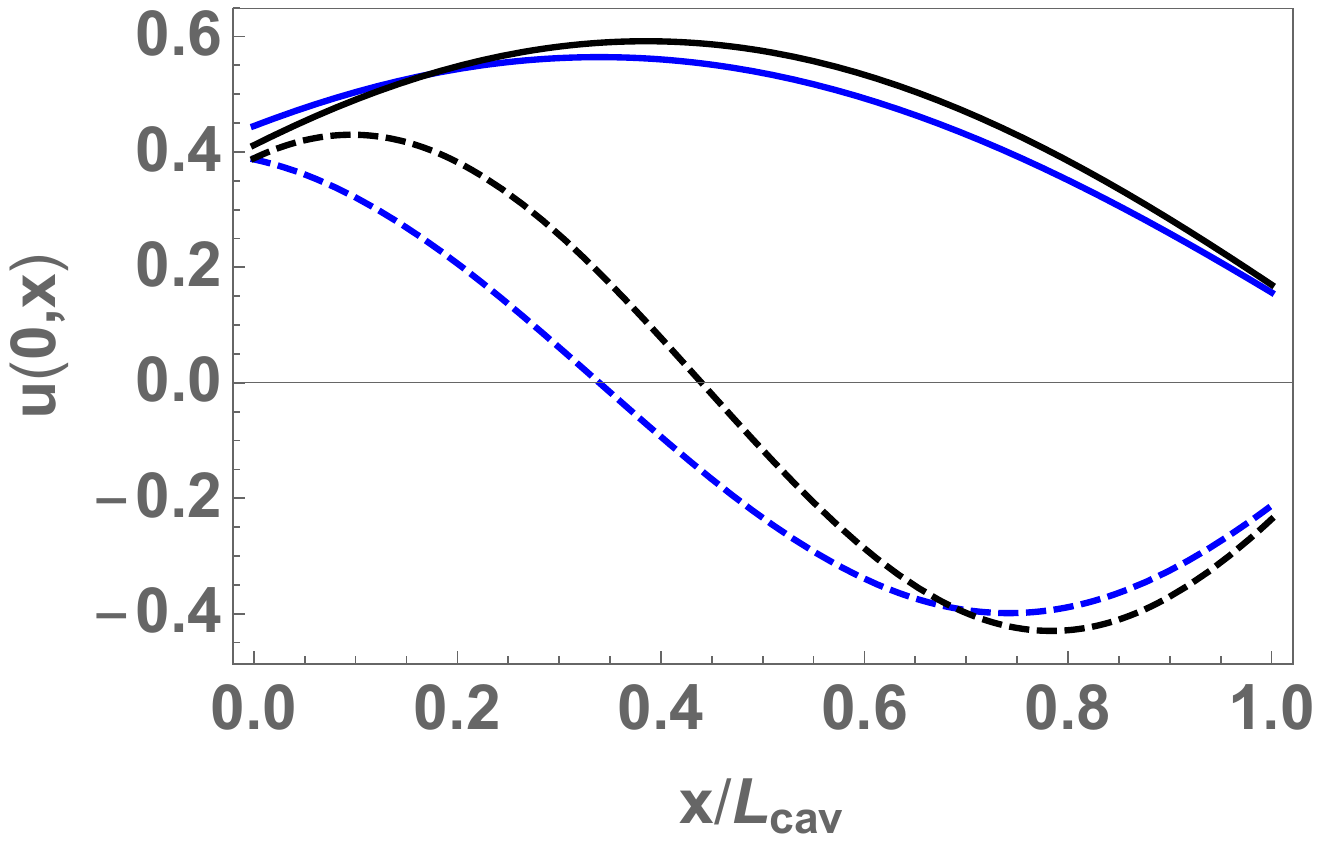}
}
\\
\subfigure{
	\includegraphics[width=0.9\columnwidth]{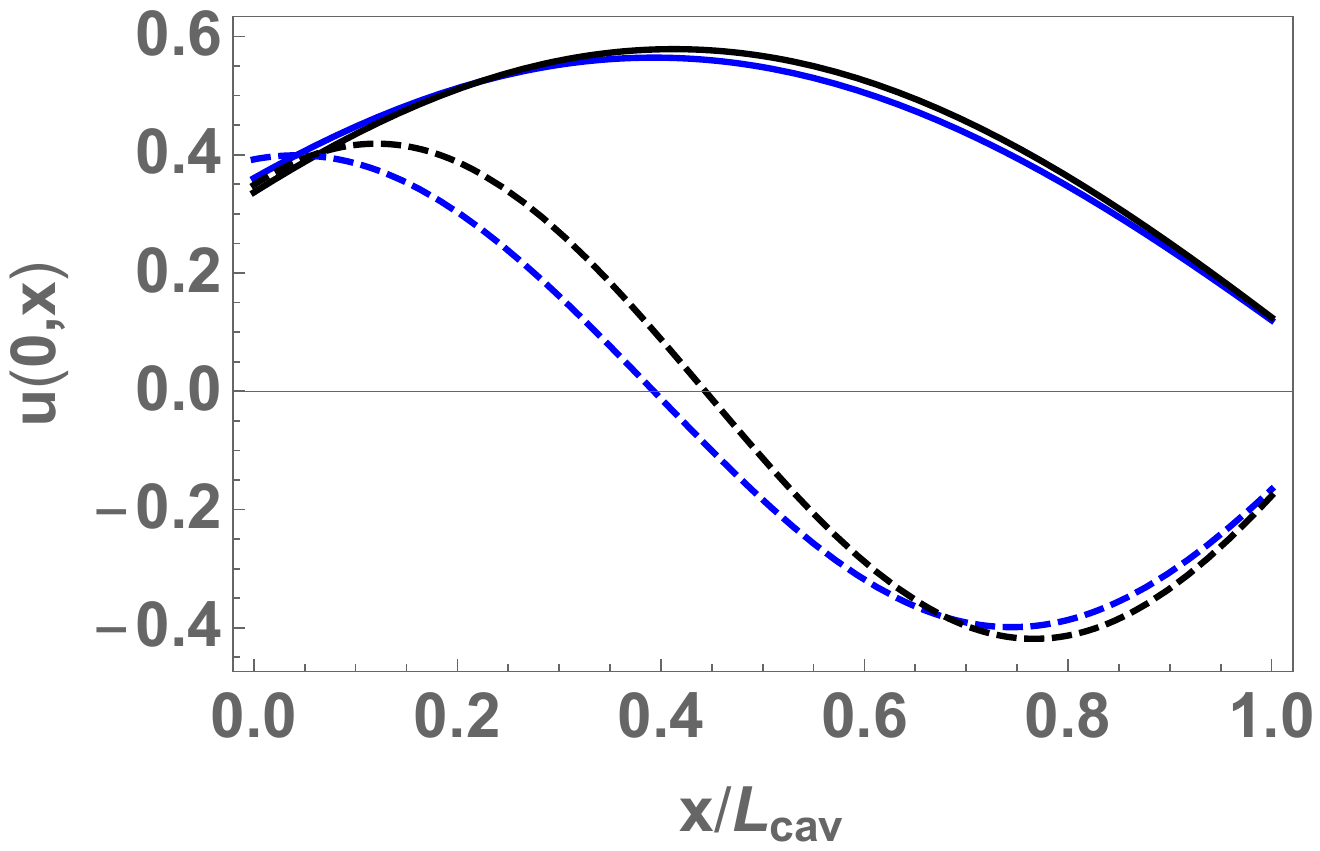}
}
\subfigure{
	\includegraphics[width=0.9\columnwidth]{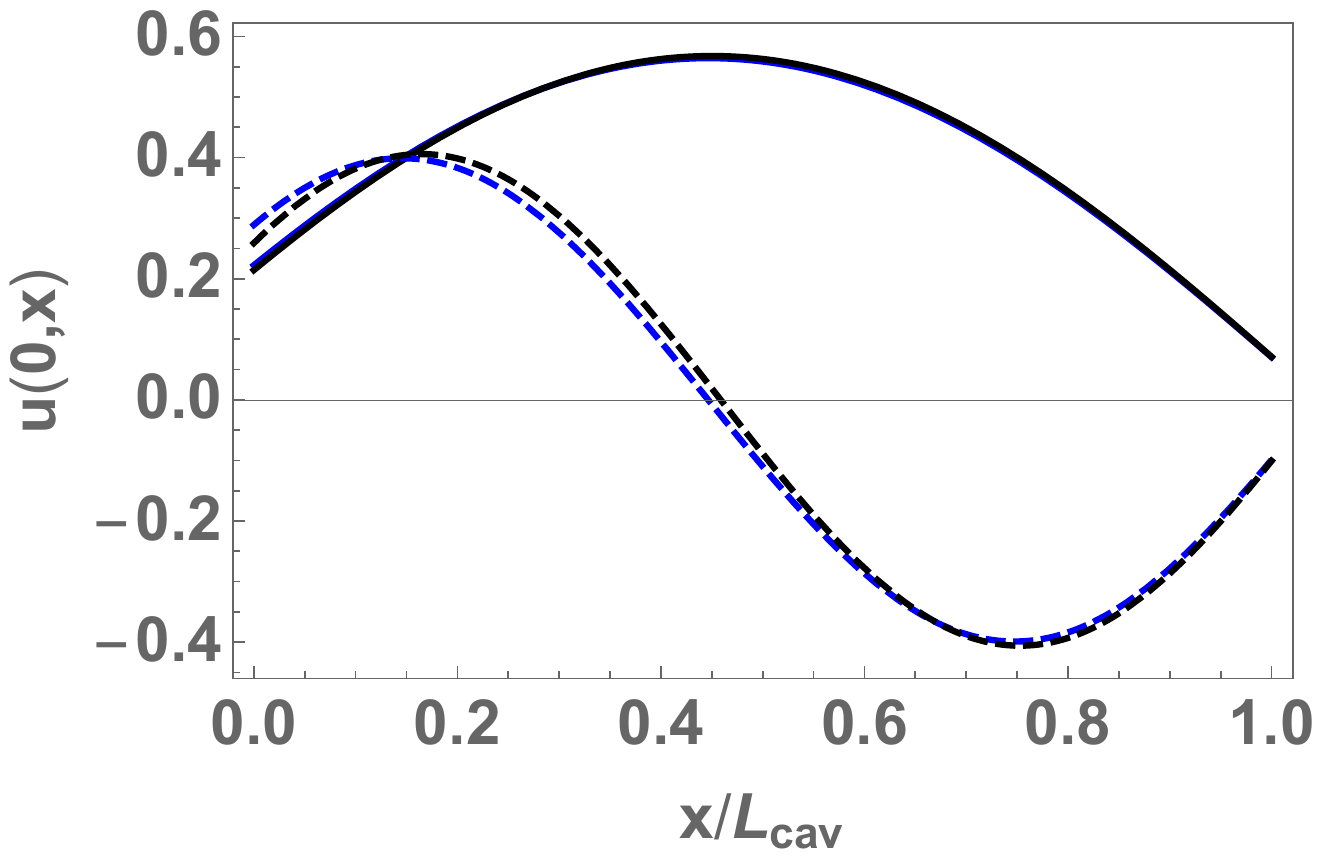}
}
\caption[]{The first two Robin mode functions \eqref{eq:robinmodefunctions} in a cavity of length $L_{\text{cav}}$ and Dirichlet mode functions \eqref{eq:dirichletmodefunctions} in a cavity of length $L=L_{\text{cav}}+d_l+d_r$. The solid blue(black) curve shows $u_1(0,x)$ in the Dirichlet(Robin) case and the dashed blue(black) curve $u_2(0,x)$ in the Dirichlet(Robin) case. In the bottom left figure, parameter values from \cite{twinparadox} are used, corresponding to $d_l/L_{\text{cav}}=0.31$ and $d_r/L_{\text{cav}}=0.096$. Top left: $d_l/L_{\text{cav}}=2\times 0.31$ and $d_r/L_{\text{cav}}=2\times 0.096$. Top right: $d_l/L_{\text{cav}}=1.5 \times 0.31$ and $d_r/L_{\text{cav}}=1.5 \times 0.096$. Bottom right: $d_l/L_{\text{cav}}=0.5 \times 0.31$ and $d_r/L_{\text{cav}}=0.5 \times 0.096$.}
\label{fig:modefunctions}
\end{figure*}
Suppose that we dynamically modulate the external fluxes during a finite time interval, with boundary conditions given by
\bea
\phi(t,0)-d_l(t)\partial_x\phi(t,x)|_{x=0}&=&0,\label{eq:robinbc12}\\
\phi(t,L_{\text{cav}})+d_r(r)\partial_x\phi(t,x)|_{x=L_{\text{cav}}}&=&0.\label{eq:robinbc22}
\eea
By computing the Bogoliubov coefficients relating the initial and final cavity modes, we can determine how the state has changed as a result of the modulation. In \cite{jormajason15}, this was done analytically for small time-dependent perturbations around static boundary conditions. With $d_{l/r}(t)=d_{l/r}^0+\delta d_{l/r}(t)$, the Bogoliubov coefficients were calculated to first order in $\delta d_{l/r}(t)$.
In this paper, we go beyond this treatment and numerically compute the coefficients for arbitrary positive time-dependent functions $d_{l/r}(t)$.

Starting with the static boundary conditions,
\bea
\phi(t,0)-d_l\partial_x\phi(t,x)|_{x=0}&=&0,\label{eq:robinbc12}\\
\phi(t,L_{\text{cav}})+d_r\partial_x\phi(t,x)|_{x=L_{\text{cav}}
}&=&0,\label{eq:robinbc22}
\eea
a complete set of normalized mode functions satisfying the Klein-Gordon equation \eqref{eq:kgminkowski} is given by
\be
u_n(t,x)=N_n\sin{\lr k_nx+\delta_n\rr}e^{-ik_nct},
\label{eq:robinmodefunctions}
\ee
where the wavenumbers $k_n$ are obtained by solving
\bea
\tan{\lr k_nL_{\text{cav}}\rr}&=&-\frac{k_n\lr d_l+d_r\rr}{1-d_ld_rk_n^2},
\eea
and the phases and normalization factors are given by
\bea
\delta_n&=&\arctan{\lr d_lk_n\rr},\\
N_n&=&\frac{1}{\sqrt{k_nL_0+\frac{\sin{\lr 2\delta_n\rr}-\sin{\lr 2(k_nL_{\text{cav}}+\delta_n)\rr}}{2}}}.
\eea

In figure \ref{fig:modefunctions}, we plot the fundamental mode of a Robin cavity of length $L_{\text{cav}}$, together with the fundamental mode of a Dirichlet cavity of length $L=L_{\text{cav}}+d_l+d_r$. For small values of $d_l/L_{\text{cav}}$ and $d_r/L_{\text{cav}}$, there is a good agreement, justifying the use of the SQUID as an analogy of a moving mirror.

In order to find the Bogoliubov coefficients of the complete evolution, we first consider a small and sudden perturbation of the boundary conditions \eqref{eq:robinbc12}-\eqref{eq:robinbc22}, resulting in a new set of static Robin boundary conditions with $d_l\rightarrow d_l+\delta d_l$ and $d_r\rightarrow d_r+\delta d_r$. Denoting the new wavenumbers, phases and normalization factors by $k'_m$, $\delta'_m$ and $N'_m$, the Bogoliubov coefficients for this instantaneous transformation are given by
\begin{widetext}
\bea
\alpha_{mn}&=&\frac{N_nN'_m}{2}\Big[\frac{k'_m+k_n}{k'_m-k_n}\lr\sin{\lr\delta'_m-\delta_n+(k'_m-k_n)L_{\text{cav}}\rr}-\sin{\lr\delta'_m-\delta_n\rr}\rr\nonumber\\
&&+\sin{\lr\delta'_m+\delta_n\rr}-\sin{\lr\delta'_m+\delta_n+(k'_m+k_n)L_{\text{cav}}\rr}\Big],\label{alpha}\\
\beta_{mn}&=&\frac{N_nN'_m}{2}\Big[\frac{k'_m-k_n}{k'_m+k_n}\lr\sin{\lr\delta'_m+\delta_n+(k'_m+k_n)L_{\text{cav}}\rr}-\sin{\lr\delta'_m+\delta_n\rr}\rr\nonumber\\
&&+\sin{\lr\delta'_m-\delta_n\rr}-\sin{\lr\delta'_m-\delta_n+(k'_m-k_n)L_{\text{cav}}\rr}\Big].\label{beta}
\eea
\end{widetext}
After time-evolving the new modes for a time $\Delta t$, a new instantaneous transformation is made. Composing transformations in this way and taking the continuum limit $\Delta t\rightarrow 0$, the Bogoliubov coefficients for a transformation with arbitrary time-dependence can be found.

\section{Simulating rigidly moving cavities}
\label{sec:simulation}
Using the method described in section \ref{sec:robin}, we can find the transformation of the cavity state for arbitrary time-dependent Robin boundary conditions. In this section, we apply this procedure to simulations of the rigid cavity trajectories described in section \ref{sec:dirichlet}. More specifically, we compute the phase shift accumulated by the state during the trajectory for both Dirichlet and Robin boundary conditions. In this section, we focus on single trajectories only, while the case of repeated trajectories is examined in section \ref{sec:resonance}.

As described in section \ref{sec:squidbc}, motion of the cavity is simulated by separately tuning the parameter $\leff(\Phi)$ for the two SQUIDs.
For a cavity moving from left to right and back, the initial value at the right SQUID should be the smallest possible one, which is $\delta L_{\text{min}}\equiv\leff(0)$. The value of $\delta L_{\text{min}}$ is set by the critical current $I_c$ of the Josephson junctions and the inductance $L_0$ per unit length of the waveguide. A smaller value means that the boundary conditions are closer to Dirichlet, but also that the effective displacement of the mirrors becomes smaller. The left mirror, on the other hand, should be initialized so that $\leff(\Phi)$ takes the maximal value used in the simulation. With the total cavity displacement \eqref{cavdisp}, this value is given by $\delta L_{\text{max}}=\delta L_{\text{min}}+d_{\text{cav}}$. To simulate a moving cavity of proper length $L$, the physical length $L_{\text{cav}}$ has to be chosen so that
\be
L=L_{\text{cav}}+\delta L_{\text{min}}+\delta L_{\text{max}}.
\ee
Since the correspondence between Robin and Dirichlet boundary conditions is only valid when the extra effective length is much smaller than the wavelength, the condition $\delta L_{\text{max}}\ll L_{\text{cav}}$ must be satisfied. This means that the cavity can only be displaced by a small fraction of its own length.

After initialization, the SQUID boundary conditions are tuned to mimic the rigid cavity mirror trajectories. Denoting the real mirror trajectories by $x_l(t)$ and $x_r(t)$, the external fluxes $\Phi_l(t)$ and $\Phi_r(t)$ through the two SQUIDs should be tuned so that
\bea
\leff(\Phi_l(t))-\delta L_{\text{max}}&=&-\lr x_l(t)-x_l(0)\rr,\\
\leff(\Phi_r(t))-\delta L_{\text{min}}&=&x_r(t)-x_r(0).
\eea

To realize the required hyperbolic waveform, it is not enough to modulate the SQUID with a single harmonic signal. Instead, an arbitrary waveform generator should be used, and the time-resolution of this device limits how short the duration of the trip segment can be made. The smaller the value of $t_a$, the larger the acceleration can be without making $d_{\text{cav}}$ too large, as can be seen in \eqref{cavdisp}.
In \cite{twinparadox}, it was assumed that $t_a=1$ ns is achievable with state-of-the-art waveform generators. As discussed in section \ref{sec:harmonicflux}, however, acceleration times as short as $t_a=0.1$ ns can also be reached. This allows us to make the value of $h$ one order of magnitude larger without losing too much precision in the Rindler-Dirichlet correspondence, see figure \ref{fig:fourier}.
\newline
\\
Let us now specifically consider the clock experiment in \cite{twinparadox} and investigate the error made when simulating the trajectories. 
The parameters used to obtain the maximal phase shift in figure 3 in \cite{twinparadox} are $t_a=1$ ns,  $L=1.1$ cm and $h = 10^{-3}$, corresponding to a cavity displacement of $d_{\text{cav}}=1.68$ mm. Moreover, we choose the physical parameters of the SQUID and the waveguide as in \cite{lindkvist6}, $L_0=0.44 \mu\mathrm{F/m}$ and $I_c=0.5 \mu\mathrm{A}$, resulting in $\delta L_{\text{min}}=0.75$ mm.  
For simplicity, let us look at a specific part of the trajectory,  when the cavity stands still in its starting position before or after the trip.  
In this case, the absolute phase shift in the cavity is proportional to the frequency of the clock mode.
Denoting the clock mode frequency of the Dirichlet (Robin) cavity by $\omega_D$ ($\omega_R$) and using the parameters specified above, we obtain $\omega_R/\omega_D = 1.56$. The correspondence can be improved if we make $\delta L_{\text{min}}$ smaller by choosing different physical parameters. For $\delta L_{\text{min}}=0.0075$ mm, we obtain $\omega_R/\omega_D = 1.48$, which is still not good enough. Thus, at least when using typical values of $I_c$ and $L_0$, we conclude that the parameter regime chosen in \cite{twinparadox} is not suitable to observe simulated time-dilation.

\begin{figure}[t]
\centering
\includegraphics[width=0.9\columnwidth]{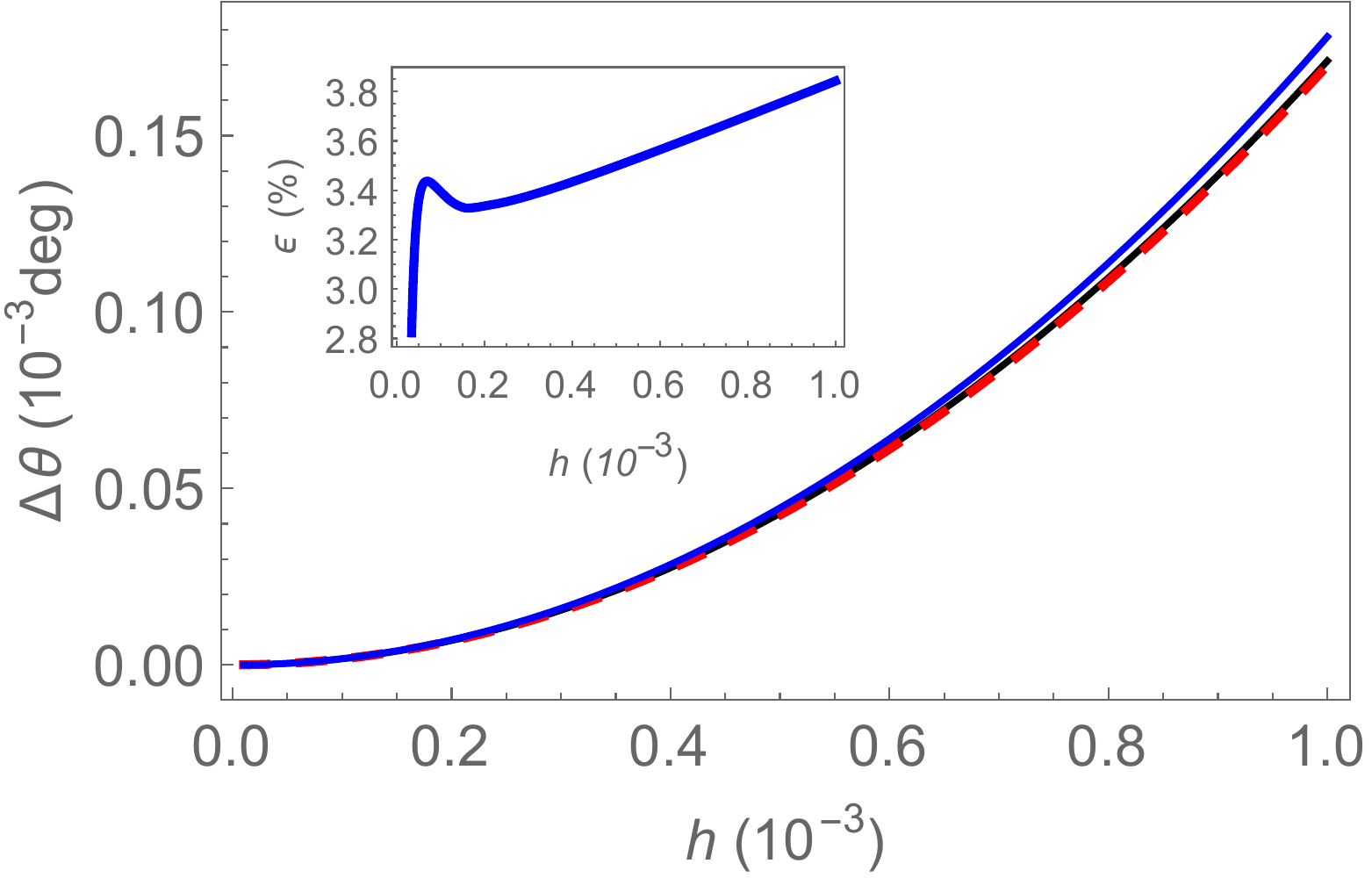}
\caption{Phase shift of a single cavity trajectory with $L =12.2$ cm, $t_a=1$ ns and $\delta_{\text{min}}=0.0075$ mm. The blue (black) solid line shows the Dirichlet (Robin) phase shift $\Delta\theta_D$ ($\Delta\theta_R$) and the red dashed line shows the time dilation for a point-like clock, scaled by the Dirichlet cavity mode frequency. The inset shows the relative error of the simulation, $\epsilon = (\Delta\theta_D-\Delta\theta_R)/\Delta\theta_D$.} 
\label{fig:basicShift}
\end{figure}

In order for the Robin cavity to better simulate a Dirichlet cavity, we can increase its length. Figure \ref{fig:basicShift} shows the phase shift of a single trajectory for a cavity with $L=12.2$ cm. Here, we see that the error is only a few percent. For 5000 trajectories, corresponding to a total travel time of 20 $\mu$s, the phase shift is of the order of 1 degree, making it measurable. Moreover, the simulated time dilation is close to that of an ideal (point-like) clock. Thus, this would be a suitable parameter regime to measure simulated time-dilation for a moving cavity clock.
\newline

In \cite{twinparadox}, the deviations of the cavity clock from the ideal clock formula were discussed. First, there is a slowing down of the clock due to its size and, second, there are additional corrections due to non-adiabatic effects, i. e. mode-mixing and particle creation. In the parameter regime suggested above, we saw that these effects are small and the total time dilation is approximately the same as for an ideal clock. 

\begin{figure}[t]
\centering
\includegraphics[width=0.9\columnwidth]{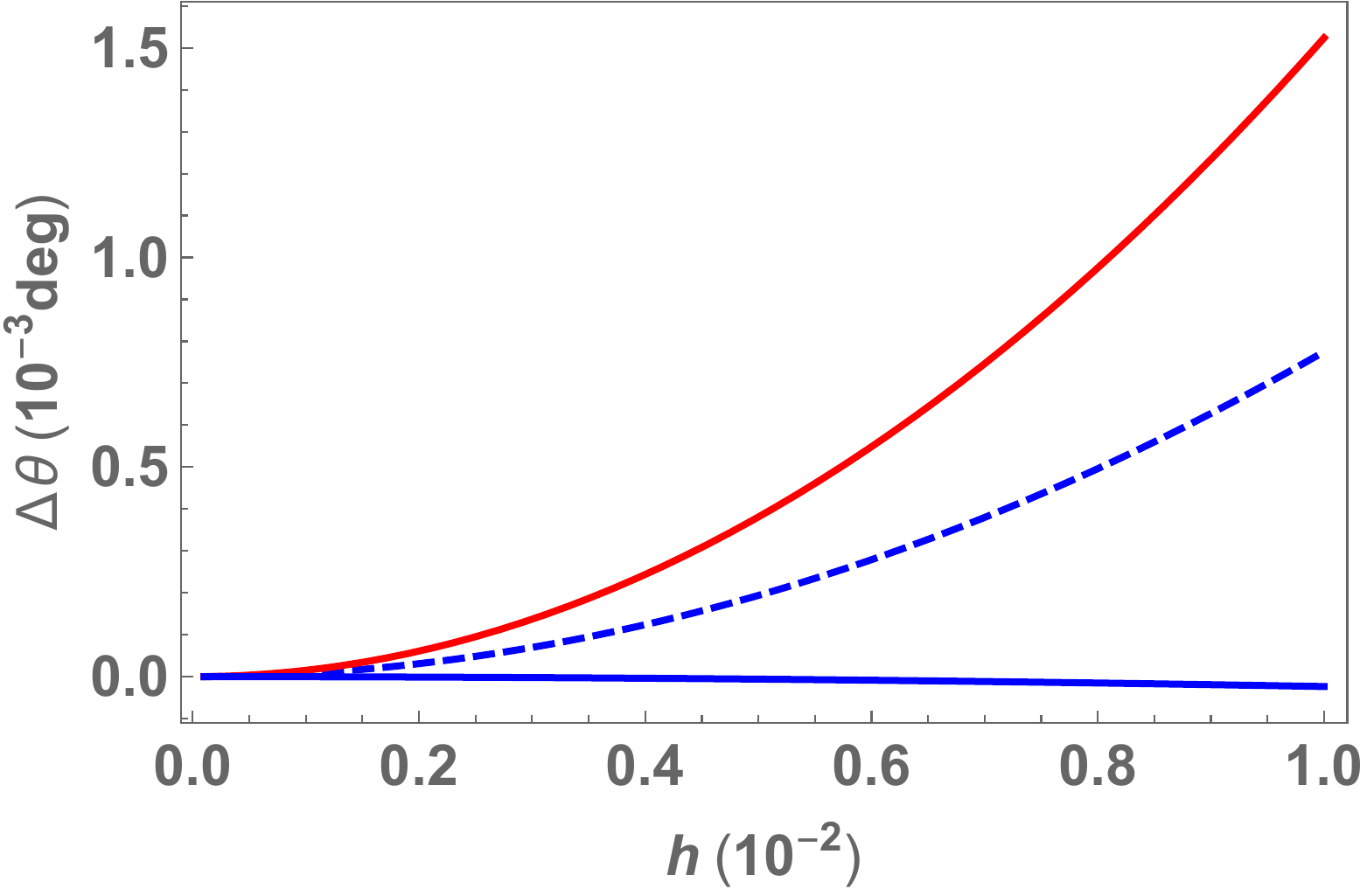}
\caption{Phase shift of a single cavity trajectory with $L =9.5$ cm, $t_a=0.1$ ns and $\delta_{\text{min}}=0.0075$ mm. The solid blue curve shows the full phase shift for a Dirichlet cavity, and the dashed blue curve the phase shift excluding the non-adiabatic effects. The red curve shows the result for a point-like clock, scaled by the Dirichlet cavity mode frequency} 
\label{fig:sizeEffects}
\end{figure}

To observe the corrections to this basic time dilation, we must explore parameter regimes corresponding to larger values of $h$. 
The two effects are difficult to decouple and in the regime where the cavity size effects are large, the non-adiabatic corrections are also comparably strong.
For example, in figure \ref{fig:sizeEffects} we plot with a dashed line the phase shift for the case where the non-adiabatic effects have been excluded (i.e. a single-mode cavity), so that, for this curve, the deviation from the ideal clock is only due to the clock size. 
Note that the slowing down of the cavity clock because of the clock size can be quite large.
Comparing to the full Dirichlet phase shift, we see that the non-adiabaticity of the transformation also greatly contributes to the final shift and a single-mode approximation generally does not suffice to describe the time evolution. The relative error in the Robin-Dirichlet approximation is in this case around $0.02\%$. 
We explore in more detail mode-mixing and particle creation in section \ref{sec:resonance}, where their effect is amplified by appropriately selecting the size of the cavity to obtain resonances for repeated trajectories.

\section{Flux tuning}
\label{sec:harmonicflux}
\begin{figure}[t]
\centering
\includegraphics[width=0.9\columnwidth]{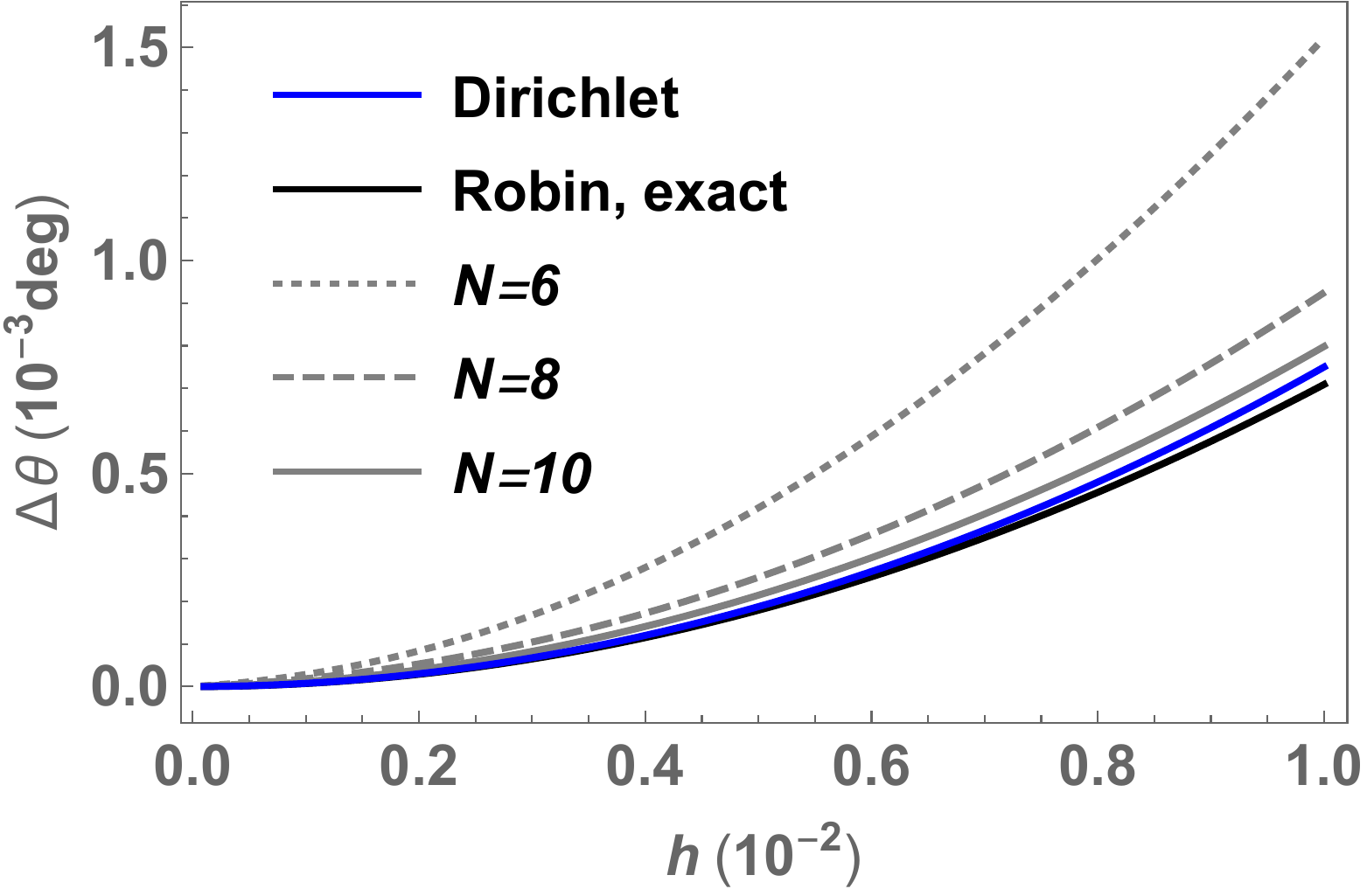}
\caption{Phase shift of a single cavity trajectory with $L =2.38$ cm, $t_a=0.1$ ns and $\delta_{\text{min}}=0.0075$ mm. The grey curves show the Robin phase shift when the trajectory has been approximated by using different numbers of Fourier harmonics.}
\label{fig:fourier}
\end{figure}

As stated in section \ref{sec:simulation}, the arbitrary waveform generators used to achieve the desired trajectories limit the frequency of the flux modulation. Tuning the flux with a harmonic signal would, on the other hand, not preserve the rigidity of the cavity. 
For a mirror trajectory of period $T=4t_a$, we can expand the flux in a truncated Fourier series as\be
\tilde{\Phi}(t)=\tilde{\Phi}_0+\sum_{n=1}^Na_n\cos{\lr\frac{2\pi n}{T}t+\delta_n\rr}.
\label{fourier}
\ee
By choosing $\tilde{\Phi}_0$, $a_n$ and $\delta_n$ separately for each SQUID as to minimize $\leff(\Phi_{l/r}(t))-\leff(\tilde{\Phi}_{l/r}(t))$, the rigid cavity trajectory can be approximated. Figure \ref{fig:fourier} shows the phase shifts calculated for these approximated trajectories for different values of $N$. We see that the result for the rigid cavity trajectory is reasonably well approximated for $N=10$. For a trajectory with $t_a=0.1$ ns, we would have $T=0.4$ ns, leading to a fundamental frequency of $2.5$ GHz. Thus, with a harmonic waveform generator capable of generating frequencies up to 25 GHz, the required waveform could be achieved for these trajectories.

\section{Resonances}
\label{sec:resonance}
As seen in figures \ref{fig:basicShift}, \ref{fig:sizeEffects} and \ref{fig:fourier}, the phase shift of a single trajectory is very small. In \cite{twinparadox}, it was suggested that the trip can be repeated many times in order to accumulate a larger phase shift. The total phase shift was calculated simply by multiplying by the number of trajectories, which is true in the regime considered there, i.e. where the non-adiabatic effects are negligible. In other regimes, however, phase dependent non-adiabatic effects become more relevant. Thus, in order to investigate the result of repeated trips, a single-mode approximation does not suffice and one has to compose several single-trip Bogoliubov transformations.
 
In section \ref{sec:simulation}, we identifed parameter regimes suitable for measuring the basic time dilation and we showed that the non-adiabatic effects are generally comparable with the corrections due to the clock size. We now show how these effects can be resonantly enhanced.
\begin{figure}[t]
\centering
\includegraphics[width=0.9\columnwidth]{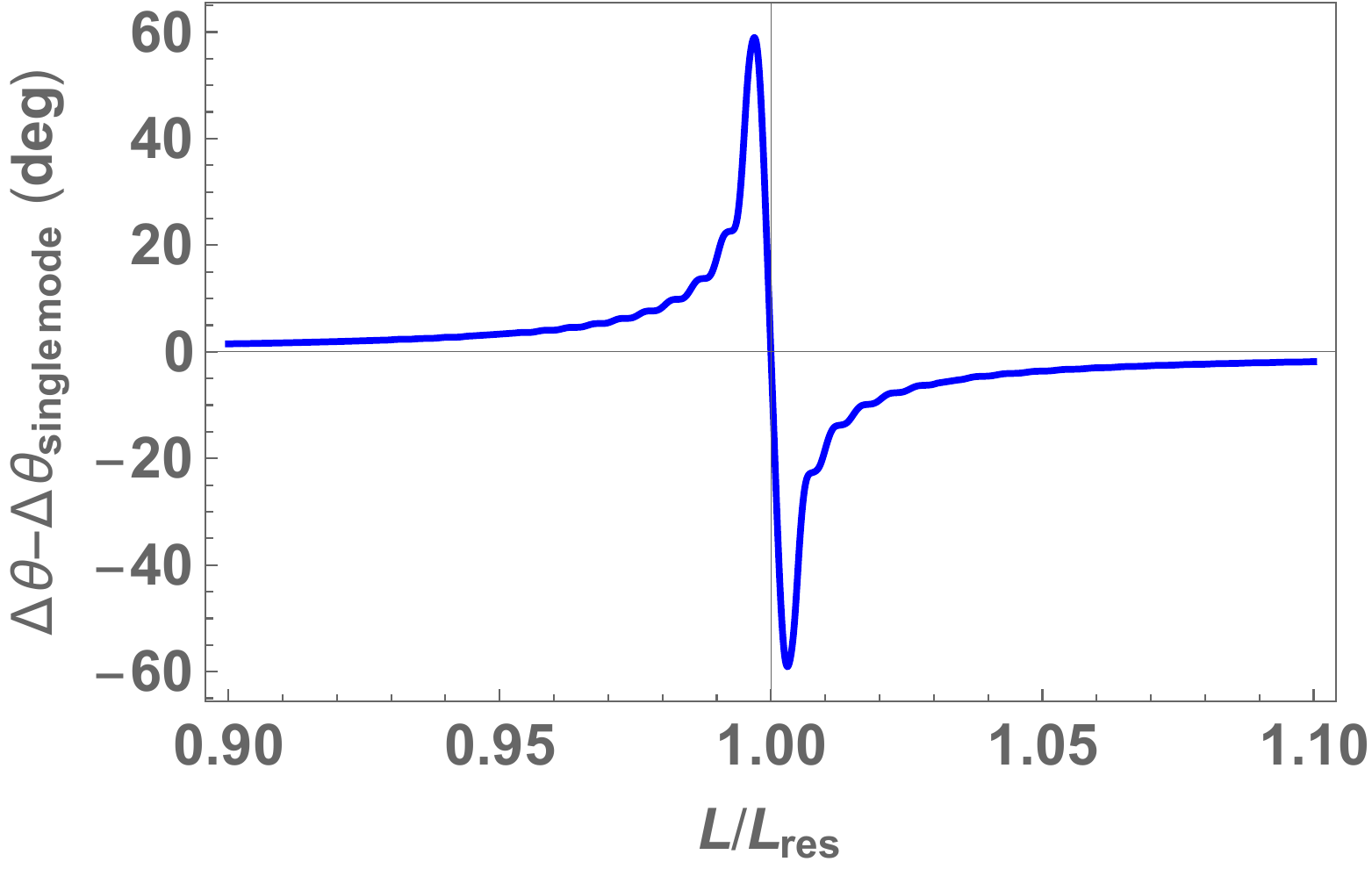}
\caption{Difference in the phase shift between the full trajectory transformation and a single-mode approximation where the non-adiabatic effects have been neglected. For the plot we considered 200 trajectories and the trajectory parameters are $t_a=0.1$ ns and $\delta_{\text{min}}=0.0075$ mm. The length at which the Bogoliubov coefficients are resonant is $L_{\text{res}}= 2.38$ cm. 
For the plot, we keep the parameter $h$ fixed to the value $h=0.85\times10^{-2}$. }
\label{fig:resonance}
\end{figure}
As discussed in \cite{reynaud96,bruschimodemixing13}, the periodic motion of the cavity leads to resonances in the Bogoliubov coefficients in \eqref{eq:bogotransf} for certain lengths. When the frequency $\omega_d$ of the trajectory matches the frequency difference $\omega_m-\omega_n$ between two modes, there is a resonance in the $\alpha_{mn}$-coefficient, accounting for mode-mixing. Likewise, when $\omega_d=\omega_m+\omega_n$, there is a resonance in the $\beta_{mn}$-coefficient, accounting for particle creation.
For the trajectories studied in \cite{twinparadox}, the Bogoliubov transformations are periodic with a period of $T=2t_a$. Thus, the condition for the particle creation resonance at $\omega_d=2\omega_1$ becomes $L = 2ct_a$. For a trajectory with $t=0.1$ ns, this is obtained by choosing $L=L_{\text{res}}=2.38$ cm. Note that for this cavity length, also the mode-mixing coefficients are in resonance.
In Fig. 7 we plot the difference in phase shift between the single-mode approximation and the full transformation for 200 trajectories as a function of the cavity proper length $L$; for the plot we keep fixed the value of the parameter $h=0.85\times 10^{-2}$.  As we can see from the figure, the non-adiabatic effects are very relevant close to the resonance, giving contributions to the phase of several degrees.

\begin{figure}[h]
\centering
\includegraphics[width=0.9\columnwidth]{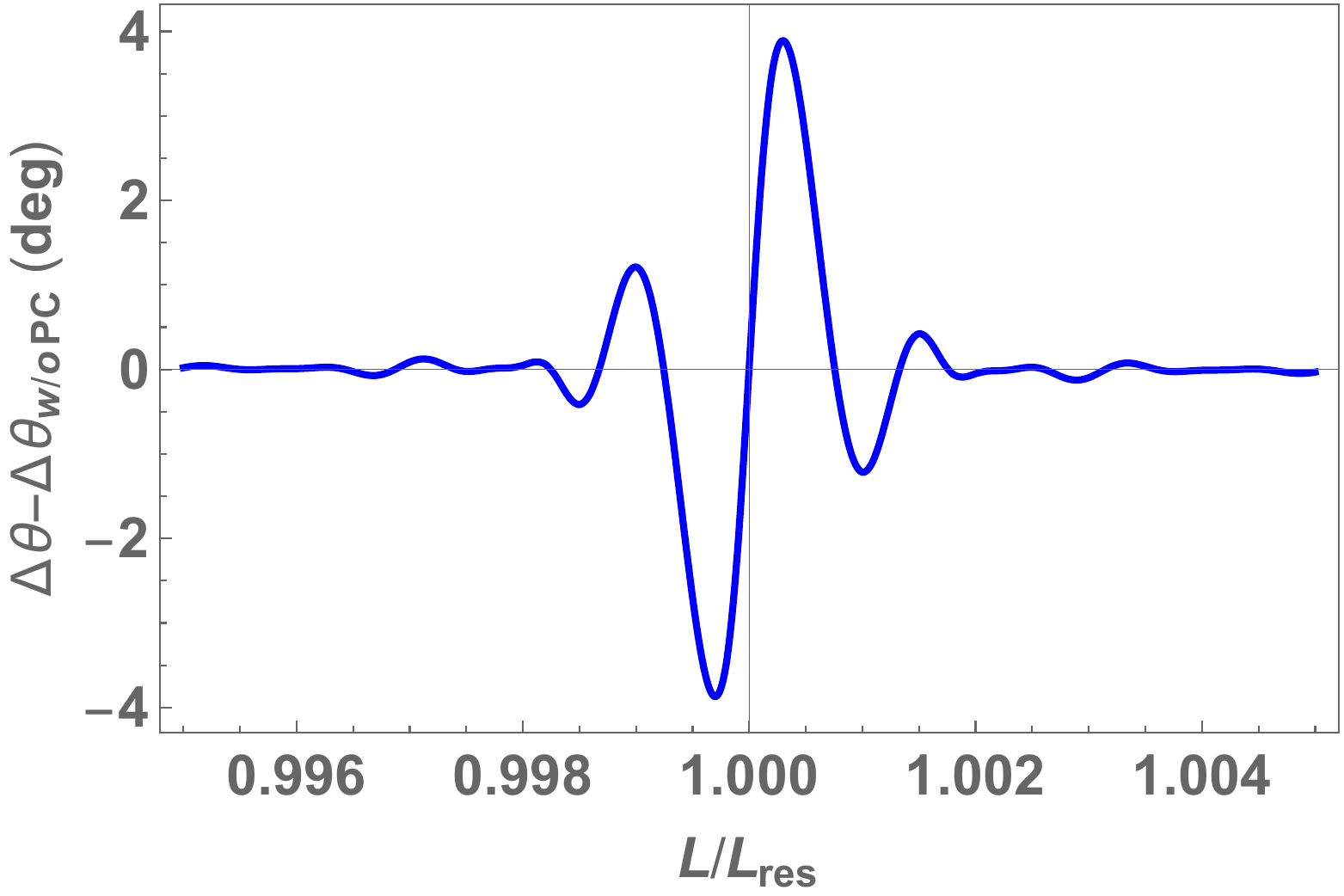}
\caption{Difference in the phase shift between the cases with and without particle creation coefficients, after 500 trips. The trajectory parameters are $t_a=0.1$ ns and $a=2\times10^{16}$ $\text{m}/\text{s}^2$ (corresponding to $h=0.34\times10^{-2}$ at resonance), and the resonance cavity length is $L_{\text{res}}=2.38$ cm.}
\label{fig:resonancezoom}
\end{figure}

In \cite{twinparadox}, the effect of the particle creation coefficients were specifically discussed.
Let us now decouple the effects of mode-mixing and particle creation and focus on the latter.
Figure \ref{fig:resonancezoom} shows the extra phase shift due to the particle creation coefficients after 500 trips. Close to resonance, the shift is large enough to be detected. It should be noted, though, that the higher modes also affect the phase shift of the clock mode close to resonance. This means that the Robin-Dirichlet correspondence is not quite as accurate as in the off-resonance case, while still reasonably good. For the parameter regimes in figures \ref{fig:resonance} and \ref{fig:resonancezoom}, the relative error is at resonance $\epsilon=4.6\%$ and $\epsilon=7.1\%$, respectively.

\section{Summary and conclusions}
\label{sec:conclusions}
In summary, we have theoretically investigated the simulation of a relativistically moving rigid cavity in superconducting circuits. Terminating a coplanar waveguide through a SQUID at each end leads to two time-dependent Robin boundary conditions for the electromagnetic field. By studying the evolution of this field and comparing it to the field in a moving rigid Dirichlet cavity, we addressed the validity of the simulation in different parameter regimes. In particular, we considered the scenario described in \cite{twinparadox}, where the cavity is used as a clock and the elapsed proper time is proportional to the phase shift. For the parameter regime suggested in \cite{twinparadox}, we found that the circuit setup does not accurately reproduce the phase shift in a mechanically moving cavity. By increasing the length of the cavity, however, we identified a different parameter regime where the correspondence holds and, at the same time, time dilation is measurable.

In addition, we examined parameter regimes where deviations of the cavity clock from an ideal clock can be measured.  First, by using larger accelerations and shorter trajectory times, we identified regimes where the slowing down of the clock due to its size and non-adiabatic effects is measurable, while the Robin-Dirichlet correspondence still holds. We also verified that a Fourier decomposition of the mirror trajectories can simulate moving rigid cavities when an appropriate number of harmonics is considered.
Because of the smallness of the output signal for single trajectories, we also explored the possibility of enhancing the phase shift by repeating the trajectories, and we found regimes where particle-creation coefficients have a measurable effect.
These findings will serve as a guideline when performing the clock experiment proposed in \cite{twinparadox}, and may also be useful when studying other effects in moving cavities like, for example, the quantum teleportation protocol proposed in \cite{teleportationfriis}.

\section*{Acknowledgments}
The authors thank the Knut and Alice Wallenberg Foundation and the Swedish Research Council for support.
\onecolumngrid 
\begin{appendix}
\section{Trajectory functions}
\label{sec:appendix}

The round-trip trajectory of an observer in the center of the cavity is, in the lab-frame coordinates $(t, x)$, given by
\be
x(t)=
\left.
  \begin{dcases}
        \sqrt{c^2t^2+c^4/a^2}, & 0\le t\le t_a  \\
        2\sqrt{c^2t_a^2+c^4/a^2}-\sqrt{\lr t-2t_a\rr^2+c^4/a^2}, & t_a\le t\le3t_a \\
	  \sqrt{c^2(t-4t_a)^2+c^4/a^2}, &3t_a\le t\le 4t_a.
    \end{dcases}
\right.
\ee
In order to preserve the rigidity of the cavity, the trajectories of the left and right mirrors must then be
\be
x_l(t)=
\left.
  \begin{dcases}
        \sqrt{c^2t^2+c^4g_-^2/a^2},&0\le t\le g_-t_a  \\
        2\sqrt{c^2t_a^2+c^4/a^2}-\sqrt{\lr t-2t_a\rr^2+c^4g_+^2/a^2},&g_-t_a\le t\le\lr 2+g_+\rr t_a \\
	  \sqrt{c^2(t-4t_a)^2+c^4g_-^2/a^2},&\lr 2+g_+\rr t_a\le t\le 4t_a
    \end{dcases}
\right.
\ee
and
\be
x_r(t)=
\left.
  \begin{dcases}
        \sqrt{c^2t^2+c^4g_+^2/a^2},&0\le t\le g_+t_a \\
        2\sqrt{c^2t_a^2+c^4/a^2}-\sqrt{\lr t-2t_a\rr^2+c^4g_-^2/a^2},&g_+t_a\le t\le\lr 2+g_-\rr t_a \\
	  \sqrt{c^2(t-4t_a)^2+c^4g_+^2/a^2},&\lr 2+g_-\rr t_a\le t\le 4t_a,
    \end{dcases}
\right.
\ee
respectively, with $g_{\pm}= (1\pm h/2).$

\end{appendix}
\twocolumngrid 
\bibliography{bibcavities}
\bibliographystyle{apsrev-nourl} 

\end{document}